\documentclass[12pt,twoside]{article} \input{epsf.sty}

\def\mypagenumber{1}
\def\mydate{Dec 18, 1998}
\def\myend{\end{document}}

\def\Journal#1#2#3#4{{#1}{\bf #2} (#3) #4}

\def\CQG{\em Class.\ Quant.\ Grav.}
\def\NPB{{\em Nucl.\ Phys.} B}
\def\PLB{{\em Phys.\ Lett.} B}
\def\PRL{\em Phys.\ Rev.\ Lett. }

\def\AP{{\em Ann.\ Phys.\ (N.Y.)} }

\def\CMP{\em Comm.\ Math.\ Phys. }

\normalsize

\newcounter{sxn}

\newcounter{axn}

\date{}

\newdimen\mybaselineskip
\newcommand{\beeq}{\begin{equation}}
\newcommand{\eneq}{\end{equation}}
\newcommand{\be}{\begin{eqnarray}}
\newcommand{\ee}{\end{eqnarray}}
\newcommand{\bpic}{\begin{picture}}
\newcommand{\epic}{\end{picture}}

\def\dd{\partial}
\def\la{\raise.16ex\hbox{$\langle$} \, }
\def\ra{\, \raise.16ex\hbox{$\rangle$} }
\def\go{\rightarrow}

\def\psibar{ \psi \kern-.65em\raise.6em\hbox{$-$} }
\def\mbar{ m \kern-.78em\raise.4em\hbox{$-$}\lower.4em\hbox{} }

\def\ep{\epsilon}

\def\n@space{\nulldelimiterspace=0pt \mathsurround=0pt }
\def\huge#1{{\hbox{$\left#1\vbox to 20.5pt{}\right.\n@space$}}}

\def\myskip{\noalign{\kern 8pt}}
\def\myeqspace{\noalign{\kern 10pt}}

\def\boxit#1{$\vcenter{\hrule\hbox{\vrule\kern3pt
    \vbox{\kern3pt\hbox{#1}\kern3pt}\kern3pt\vrule}\hrule}$}
\def\bigbox#1{$\vcenter{\hrule\hbox{\vrule\kern5pt
     \vbox{\kern5pt\hbox{#1}\kern5pt}\kern5pt\vrule}\hrule}$}

\def\ignore#1{{}}

\begin{document}

\bibliographystyle{unsrt}
\footskip 1.0cm

\thispagestyle{empty}
\setcounter{page}{\mypagenumber}

             
\begin{flushright}{\mydate ~ (.)
\\  UMN-TH-1732/98  , TPI-MINN-98/28-T \\
}
\end{flushright}

\vspace{2.5cm}
\begin{center}
{\LARGE \bf {Monopole-Instanton Type Solutions  }}\\
\vskip .8cm 
{\LARGE \bf {In 3D Gravity}}\\
\vspace{3cm}
{\large Bayram Tekin\footnote{e-mail:~ tekin@mnhepw.hep.umn.edu} }\\

\vspace{.5cm}
{\it School of Physics and Astronomy, University of Minnesota}\\ 
{\it  Minneapolis, MN 55455, U.S.A.}\\ 
\end{center}

\vspace*{2.5cm}


\begin{abstract}
\baselineskip=18pt
Three dimensional Euclidean gravity in the dreibein-spin connection
formalism is investigated. We use the monopole-instanton ansatz for
the dreibein and the spin connection. The equations of motion are 
solved. We point out a two dimensional solution 
with a vanishing action.

\end{abstract}
\vfill



 
\newpage



\normalsize
\baselineskip=22pt plus 1pt minus 1pt
\parindent=25pt

2+1  dimensional gravity is an interesting model which carries some
features of the 3+1 dimensional general relativity. This model
has been studied in the literature in great detail with many different
approaches \cite{Deser1,'t Hooft,Deser2,Townsend,Witten,Carlip} .
Although there are no physical degrees of freedom 
and there is not a proper 
Newtonian limit \cite{Barrow} , i.e. two static masses experience no
force, three dimensional gravity is still far from being trivial.
Ba\~{n}ados,
Teitelboim and Zanelli
showed that there exists a solution which looks like the 3+1 dimensional
black hole \cite{BTZ}. BTZ  black hole has an event horizon and its
thermodynamic properties resemble those of the ``realistic'' black
hole. Unlike the four dimensional case, three dimensional black hole 
can not have a curvature singularity at the
origin. But one can still study many classical and quantum properties of
the 3+1 dimensional black hole through the BZT solution.

Pure gravity
(with or
without cosmological
constant) was shown to be equivalent to Chern-Simons theory
\cite{Townsend,Witten}. This  equivalence led Witten \cite{Witten} to
prove that the theory is quantizable and renormalizable.  
In the
Ach\'{u}carro-Townsend-Witten formulation  one combines the
dreibein and the spin connection into a single gauge field
whose dynamics is
governed by the Chern-Simons Lagrangian.   
 
Conformal Weyl tensor vanishes in 2+1 dimensions and the Riemann tensor
is determined by the Ricci tensor. The solutions to the Einstein's
equations consist of flat spaces. 
In this letter we will be interested in the spaces with
vanishing cosmological constant.
The existence of non-trivial global geometry makes flat
spaces interesting to study.

In the path integral formulation of gravity , in the saddle point
approximation, one is interested in the finite action solutions to the
Euclidean equations of motion.
Our purpose is to find monopole-instanton type solutions
for Euclidean gravity with zero  cosmological constant.
We use the monopole ansatz for
both the spin connection and the dreibein which are assumed to be
independent fields. In the metric formulation of gravity 
metric has to be non-degenerate.
In the dreibein-spin connection
formalism dreibein need not be invertible. So one can obtain
degenerate metrics. We adopt the later formalism.
The results of our work depend on this
crucial difference. We would like to stress that the non-invertibility
of the dreibein is not unnatural if the spin connection and the dreibein
are
independent fields \cite{Witten,D'Auria}. In fact Witten \cite{Witten}
shows
that
to make sense of the quantum theory of gravity one needs non-invertible
dreibeins.

In three dimensions the Einstein-Hilbert action can be written
the following
form 
\be
S_{EH} = {1\over{16\pi G}} \int_{\cal{M}} d^3x \epsilon^{ijk}\,
\eta_{ab}\, e^a\,_i \, R^b\,
_{kj} 
\label{EH}
\ee
where $e^a\,_i$ is the dreibein and $R^a\, _{ij}$ is the Riemann tensor. 
The indices $(a,b,c)$ denote the local Lorentz frame and
$(i,j,k)$ are the non-inertial frame indices. The metrics , $\eta_{ab}$
and
$g_{ij}$ have Euclidean signature. 
Explicitly the Riemann tensor is 
\be
R^a\,_{kj} = \dd_k w^a\,_j -\dd_j w^a\,_k +\epsilon^a\,_{bc}w^b\,_k 
w^c\,_j
\label{Rieman}
\ee
We have used the fact that in three dimensions one can treat the 
dreibein and the dual of the spin connection as the fundamental fields.
The dual of the spin connection is a one form defined 
through $ w^a = {1\over 2} \epsilon^{abc}w_{bc}$. 

The symmetry group of the theory is $GL(3,R) \times SO(3)$. 
The first
factor refers to the general coordinate transformations ,i.e. $x\go x'(x)$  
which are realized by the fields in the following way. 
\be
e^{a'}\,_i(x') = {\dd x^j\over \dd x'^i}
e^a\,_j(x) \hskip 2 cm
w^{a'}\,_i(x') ={\dd x^j\over \dd x'^i} w^a\,_j(x) 
\ee
$SO(3)$ is the group of local Lorentz rotations under which the
fields transform in the following way. 
\be
e^{a'}\,_i(x) = \Lambda^a\,_b(x) e^b\,_i(x) \hskip 2 cm
w^{a'}\,_i(x) = \Lambda^a\,_b(x) w^b\,_i(x)
\ee 
$\Lambda$ is an element of $SO(3)$.
 It is clear that $R^a\,_{kj}$ transform as a tensor under $GL(3,R)$
and as a vector under $SO(3)$.
One
should observe that unlike the case of non-Abelian Chern-Simons theory
the constant multiplying  the action is not quantized. $G$ has
the dimension of length. 

At the classical level the main question is to find the manifolds which
minimize the action $S_{EH}$.   
Equations of motion can be obtained by varying the action with respect to 
the dreibein and the spin connection. In doing so one obtains

\be
&&\epsilon^{ijk}\dd_i e^a\,_j + \epsilon^{ijk}\,\epsilon^a\,_{bc}\,e^b\,_i
w^c\,_j=0 \nonumber \\
&&\epsilon^{ijk}R^a\,_{jk}= 0
\label{eqnsofmotion}
\ee
The first equation is the torsion free condition and the second equation
is analogous to Einstein equation in four dimensions. Since there are no
matter fields  and the cosmological constant is zero the 3D universe is
flat. 
We would like to
point out that invertibility of the dreibein is not required. 

We are interested in the solutions where the dreibein and the spin
connection are of the  monopole type.
\be
&&e^a\,_j(\vec{x})= {G\over r} \left[ -\epsilon^a\,_{jk}\,
\hat{x}^k\,\phi_1 + \delta^a\,_j\,\phi_2 +(r A-\phi_2)\,\hat{x}^a
\hat{x}_j \right]\\          
&&w^a\,_j(\vec{x})= {1\over r} \left[ \epsilon^a\,_{jk}\,
\hat{x}^k\,(1-\psi_1)+ \delta^a\,_j\psi_2 +(r B-\psi_2)\hat{x}^a
\hat{x}_j\right]
\label{ansatz}
\ee
The functions $A$, $B$, $\phi_i$ and $\psi_i$ depend on $r$ only. Writing
the first term in the dreibein as above simplifies the action and 
the equations of
motion. The constant
$G$ is included in the  definition of the dreibein to keep the dreibein 
dimensionless.  It is clear that the monopole ansatz breaks
$GL(3,R)\times SO(3)$ to $SO(3)$. 
Inserting the spin connection in equation (\ref{Rieman}) one obtains the
Riemann tensor for the monopole ansatz. 
\be
R^a\,_{ij} &=&
 {1\over r^2}
  \ep_{ij b}\,  \hat x^a \hat x^b\, (\psi_1^2 + \psi_2^2 -1)
+ {1\over r} (\ep^a\,_{ij} -  \ep_{ij b}  \hat x^a \hat x^b )
(\psi_1' + B\psi_2) \cr
\noalign{\kern 10pt}
&& \hskip 4cm + (\delta^a\,_j \hat x_i - \delta^a\,_i \hat x_j) 
{1\over r}(\psi_2' - B\psi_1)
\label{Riemann tensor}
\ee
Prime denotes the derivative. The metric $g_{ij}$ can be recovered through
the relation 
$ g_{ij} = \eta_{ab} e^a\,_i e^b\,_j$ which yields; 
\be
g_{ij} = {G^2\over r^2}\Bigg\{ (\phi_1^2 + \phi_2^2)
(\delta_{ij} -\hat x_i \hat x_j)
         + r^2 A^2 \hat x_i \hat x_j\Bigg\}   
\ee
$A$ and $\phi_i$ are to be
determined
from the equations of motion.
The metric is degenerate if $A=0$. It is easy to show that in general one
can not
bring $A$ to zero by local Lorentz or general coordinate
transformations.

Making use of the equations (6) and (8) the Einstein-Hilbert action
reduces to the following form. 
\be
S = - \int_0^\infty dr \, \Bigg\{
 \psi_a' \epsilon_{ab}\phi_b +B\psi_a \phi_a
+ {A\over 2}(\psi_a \psi_a -1) \Bigg \}
\label{action}
\ee
where $\{a,b\}=(1,2)$ and $ \epsilon_{ab}$ is antisymmetric. Summations
are
implied over the repeated indices.
The equations of motion can be found either through (\ref{eqnsofmotion})
or by varying (\ref{action}) with
respect to the six fields.
\be
\label{eqn1}
\epsilon_{ab}\phi_b' - B\phi_a - A\psi_a = 0 \\
\label{eqn2}
\epsilon_{ab}\psi_b' - B\psi_a  = 0 \\
\label{eqn3}
\phi_a \psi_a = 0 , \hskip 1.5 cm \psi_a \psi_a = 1
\ee
This system can be solved easily. The last line states that $\psi_a$
is a unit two-vector and $\phi_a$ is orthogonal to  $\psi_a$.
So we can set $\phi_a = \epsilon_{ab}\psi_b f(r)$, where $f(r)$ is an
arbitrary function. Equation (\ref{eqn1})
gives $A= - f'(r)$. Setting $\psi_1 = \cos\Omega(r)$ and 
$\psi_2 = \sin\Omega(r)$ one obtains $B =\Omega'(r)$ from equation
(\ref{eqn2}). The regularity of the spin connection at the
origin requires $\Omega(0)=0$. The following is the summary of our
solution
\be
&&\psi_1 = \cos\Omega(r) , \hskip 2.5 cm \psi_2 = \sin\Omega(r), \hskip
3cm  
B =\Omega(r)' \nonumber \\ 
&&\phi_1 = f(r)\sin\Omega(r), \hskip 1.8cm \phi_2 = -
f(r)\cos\Omega(r), 
\hskip 1.9cm A= - f(r)'
\ee 
The two functions , $f(r)$ and $\Omega(r)$ , can not be determined from
the equations of motion of course. Both of them represent the gauge
degrees of
freedom in the theory.
The metric becomes
\be
g_{ij} =  G^2\Bigg\{(\delta_{ij} - \hat{x}_i\hat{x}_j)
{f^2\over r^2} + f'\,^2\hat{x}_i\hat{x}_j \Bigg\}
\label{metric}
\ee
Unlike $\Omega(r)$, we have not imposed any boundary condition on $f(r)$.
If $f(r)= r/G$ one obtains
$g_{ij}= \delta_{ij}$ , which is the trivial Euclidean space. If
$f(r)= 1 $ then  $g_{ij} =  G^2 (\delta_{ij} - \hat{x}_i\hat{x}_j)/r^2$.
In the polar coordinates this solution is a two sphere;
${ds}^2= G^2 (d\theta^2 +\sin^2\theta d\varphi^2)$.
Choosing $f(r)$ to be a constant is possible only because there is no
regularity condition on the dreibein at the origin. This choice leads to
a singular dreibein since $\phi_2$ does not vanish at $r=0$.
In the metric formulation 
one can start with equation (\ref{metric}) but one can not choose $f(r)$
to
be a constant if one wishes to write down an action.  

In general if $f(r)$  does not vanish at the origin the dreibein is
singular. But among those singular dreibeins the only dreibein
that leads to a two dimensional
space is when $f(r) = C $  as mentioned above. The radius of the two
dimensional sphere is not determined since any constant $C$ is fine.   

For completeness let us write the dreibein and the spin connection for
our solution.
\be 
&&e^a\,_j(\vec{x})= - {G\over r} \left[\epsilon^a\,_{jk}\,
\hat{x}^k\,f\sin\Omega + \delta^a\,_j\,f\cos\Omega +(r
f'-f\cos\Omega)\,\hat{x}^a
\hat{x}_j \right]\\
&&w^a\,_j(\vec{x})= {1\over r} \left[ \epsilon^a\,_{jk}\,
\hat{x}^k\,(1-\cos\Omega)+ \delta^a\,_j\sin\Omega +(r\Omega' 
-\sin\Omega)\hat{x}^a\hat{x}_j\right]
\ee
The action for this solution vanishes as expected. We have omitted 
a surface term and/or  gauge fixing terms since our analysis has been 
classical. In the quantum theory one has to impose gauge fixing 
conditions on the dreibein and the spin connection.  

In this letter we have made use of the monopole-instanton ansatz for the
spin connection and the dreibein. At the classical level, one learns 
from this analysis that there are two dimensional solutions in this
theory with finite (in fact zero) action.
In principle it is possible to develope a quantum theory near
these solutions. In the metric formulation this is not possible.
Although spin connection is usually introduced to couple gravity
to spinors, in the above analysis its existence is the reason that
we can have two dimensional solutions.

Some future directions of the research would be to include a
cosmological constant and matter in this theory. Another interesting
problem is
to add the Chern-Simons gravity \cite{Templeton} term in the action 
and study the monopole-instanton solutions. In the Euclidean space
Chern-Simons term might be imaginary (depending on how one
defines the Wick rotation). In this case one would be looking at 
the complex-monopole type dreibeins and  spin-connections in
general. Complex monopoles were introduced in \cite{Tekin} in the context
of a gauge theory.

\leftline{\bf Acknowledgements}
I am grateful to  Yutaka Hosotani for his encouragement,
for many enlightening discussions and for his critical reading of the
manuscript. I would like to thank Arkady Vainshtein for valuable
suggestions. Roman Jackiw , whose comments led to a revision of
the paper, is acknowledged with great appreciation. 
This work was supported by the
Graduate School of University of Minnesota through the 
Dissertation Fellowship.  

\vskip 1cm

\leftline{\bf References}  

\renewenvironment{thebibliography}[1]
        {\begin{list}{[$\,$\arabic{enumi}$\,$]}  
        {\usecounter{enumi}\setlength{\parsep}{0pt}
         \setlength{\itemsep}{0pt}  \renewcommand{\baselinestretch}{1.2}
         \settowidth
        {\labelwidth}{#1 ~ ~}\sloppy}}{\end{list}}

\myend
\begin{thebibliography}{99}

\small

\bibitem{Deser1}
S.\ Deser ,R.\ Jackiw and G.\ 't Hooft ,\Journal {\AP}{152}{1984}{220}

\bibitem{'t Hooft}
G.\ 't Hooft ,\Journal {\CMP}{117}{1988}{685}

\bibitem{Deser2}
S.\ Deser and R.\ Jackiw  ,\Journal {\CMP}{118}{1988}{495}

\bibitem{Townsend}
A.\ Achuc\'{a}rro and P.K.\ Townsend, 
\Journal{\PLB}{180}{1986}{89}

\bibitem{Witten} 
E.\ Witten, \Journal{\NPB}{311}{1988}{46}

\bibitem{Carlip}
S.\ Carlip, preprint gr-qc/9503024 ; M.\ Welling preprint hep-th/9511211;
H-J.\ Matschull preprint gr-qc/9506069

\bibitem{Barrow}
J.D.\ Barrow ,A.B.\ Burd and D.\ Lancaster ,\Journal {\CQG}{3}{1986}{551}

\bibitem{BTZ}
M.\ Ba\~{n}ados, C.\ Teitelboim and J.\ Zanelli,
\Journal{\PRL}{69}{1992}{1849} 

\bibitem{D'Auria} 
R.\ D'Auria and T.\ Regge , \Journal{\NPB}{195}{1982}{308}


\bibitem{Templeton}
S.\ Deser , R.\ Jackiw and S.\ Templeton ,\Journal{\PRL}{48}{1982}{976} 


\bibitem{Tekin}
B.\ Tekin , K.\ Saririan and Y.\ Hosotani preprint hep-th/9808045 ;
hep-th/9808105



\end{thebibliography}
